# A Novel CMAQ-CNN Hybrid Model to Forecast Hourly Surface-Ozone Concentrations Fourteen Days in Advance

Authors: Alqamah Sayeed[a], Yunsoo Choi[*,a], Ebrahim Eslami[a,b], Jia Jung[a], Yannic Lops[a], Ahmed Khan Salman[a]
[a]Department of Earth and Atmospheric Sciences, University of Houston, TX 77004
[b]Houston Advanced Research Center, The Woodlands, TX 77381
*corresponding author: ychoi23@central.uh.edu

---

Issues regarding air quality and related health concerns have prompted this study, which develops an accurate and computationally fast, efficient hybrid modeling system that combines numerical modeling and machine learning for forecasting concentrations of surface ozone. Currently available numerical modeling systems for air quality predictions (e.g., CMAQ, NCEP EMP) can forecast 24 to 48 hours in advance. In this study, we develop a modeling system based on a convolutional neural network (CNN) model that is not only fast but covers a temporal period of two weeks with a resolution as small as a single hour for 255 stations. The CNN model uses forecasted meteorology from the Weather Research and Forecasting model (processed by the Meteorology-Chemistry Interface Processor), forecasted air quality from the Community Multi-scale Air Quality Model (CMAQ), and previous 24-hour concentrations of various measurable air quality parameters as inputs and predicts the following 14-day hourly surface ozone concentrations. The model achieves an average accuracy of 0.91 in terms of the index of agreement for the first day and 0.78 for the fourteenth day while the average index of agreement for one day ahead prediction from the CMAQ is 0.77. Through this study, we intend to amalgamate the best features of numerical modeling (i.e., fine spatial resolution) and a deep neural network (i.e., computation speed and accuracy) to achieve more accurate spatio-temporal predictions of hourly ozone concentrations. Although the primary purpose of this study is the prediction of hourly ozone concentrations, the system can be extended to various other pollutants.

## Main

Surface ozone can pose a significant health risk to both humans and animals alike, and it also affects crop yields (USEPA - 2006). According to the US Clean Air Act, it is one of the six most common air pollutants and considering its impact on health, the Environmental Protection Agency (EPA) of the United States has limited the maximum daily eight-hour average (MDA8) concentration of ozone to 70 ppb. Similarly, the Ministry of Environment in South Korea has declared a standard for hourly ozone of 100 ppb and 60 ppb for MDA8. To achieve these attainment goals and to understand future projections (forecasts), researchers have turned to various numerical modeling and statistical analysis tools. One such numerical model is the Community Multi-scale Air Quality Model (CMAQ), a chemical transport model (CTM) developed by the USEPA[1]. Widely used to forecast the air quality of a region with considerable accuracy, CMAQ is an open-source multi-dimensional model that provides estimated concentrations of air pollutants (e.g., ozone, particulates, $NO_x$) at fine temporal and spatial resolutions. It has been used as a primary dynamical model in regional air pollution studies; CMAQ modeling, however, has several limitations (e.g., parameterization, simplified physics, and

chemistry) and raises uncertainties that lead to significant overestimations of ozone concentrations.[2–5]

CTMs require substantial computational time since they entail multiple physical processes (e.g., transport, advection, deposition, and chemistry) for each grid. The fastest compilation time is 33 minutes.[6] Unlike CTMs, machine learning(ML) can be trained to forecast multi-hour output using a certain set of inputs more accurately within faster processing time.[7,8] In addition, it requires only one training process, further reducing the computational time. Although all ML models are more accurate with faster processing speeds, they are very localized (station-specific) and generate large underpredictions of daily maximum ozone concentrations.[7,9,10]

The objective of using this ML technique is to enhance the CMAQ modeling results by taking advantage of i) the deep neural network (DNN), a computationally efficient, artificially intelligent system that recognizes uncertainties resulting from simplified physics and chemistry (e.g., parameterizations) of the CMAQ model; and ii) CMAQ, which computes unmeasured chemical variables along with fine temporal and spatial resolutions. The aim of this approach is to use the best of both numerical modeling and ML to design a robust and stable algorithm that more accurately forecasts hourly ozone concentrations 14 days in advance and covers a larger spatial domain.

## Discussion

We trained the models based on two loss functions (methods 1 and 2) and fourteen days (28 different models), from January 1, 2014, 0000UTC to December 31, 2016, 2300UTC. After training the models, we evaluated them based on various performance parameters. The models based on both methods reported the highest IOA for prediction one-day ahead, but the IOA decreased on subsequent days. The average IOAs (method 1 – 0.90, method 2 – 0.91) and correlations (method 1 – 0.82, method 2 – 0.83) for one-day ahead prediction were comparable. The performance of both methods showed improvement over that of the CMAQ model (IOA-0.77, correlation-0.63). The IOA of method 1 increased by 16.86% and that of method 2 by 17.98%. The correlation of method 1 increased by 30% and that of method 2 by 32%.

**Performance Comparisons of CMAQ and CNN models**
Figure 1 shows the yearly IOA (average of all stations). The IOA decreased sharply from day 1 to day 3 but stabilized after the three-day forecasts from both methods. The IOA for day 4 was lowest during the first week of prediction for method 1. After day 4, the IOA increased until day 6 and then decreased until day 10. It increased slightly on day 11 but then decreased further. For method 2, the IOA decreased until day 5, increased until day 7, and then further decreased after day 8. One possible explanation for the weekly trend relates to the weekly cycle of ozone concentrations.[11] That is, observed ozone followed a weekly cycle, exhibiting a decreasing trend in its correlation until day 3 and then an increasing trend until it peaked on day 7. The same cycle occurred during the second week. Also, the figure depicts the superior performance of method 2 to that of method 1. The average increase in the IOA of method 2 compared to that of method 1 was 4.77%; a maximum increase of 6.64% occurred on day 4, and a minimum increase of less than 1 % occurred on day 1. The greatest increase in the IOA happened on the worst-performing days (days 4, 13,8, 7, and 12 show an increase of 6.6, 5.8, 5.6, 5.4, and 5.3%, respectively) by method 1.

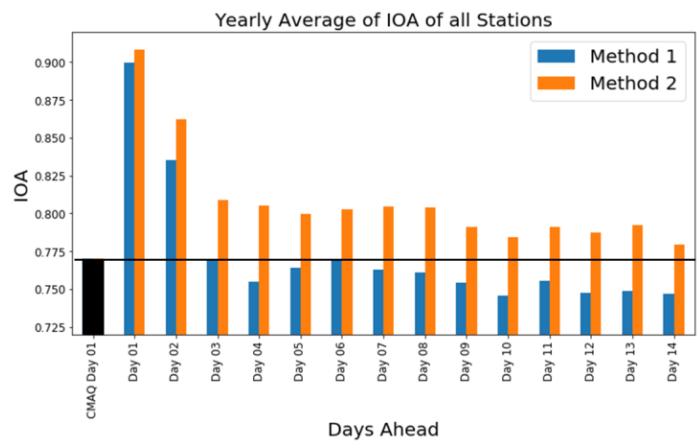

**Figure 1:** Comparison of Index of Agreement for two-advance prediction using Method 1 and 2. x-axis in the plot shows the days ahead, and the y-axis represents the index of agreement. The blue line represents IOA of each day advance prediction using Method 1 (mean squared error as loss function). The orange line represents IOA of each day advance prediction using Method 2 (Index of Agreement as loss function).

**Performance Evaluation of Selected Method**

It is evident from the above discussion that the performance of method 2 overshadowed that of method 1; therefore, we further analyze the performance of method 2 below. Figure 2 lists the average yearly IOA of each district in South Korea. If a district had more than one station, we averaged its IOAs. We found that inland cities performed slightly better than the coastal ones, and their performance improved the farther they were from the coast (Figures 2 & 3 and Figure S3 in the supplementary document). For example, Seoul performed slightly better than Incheon, the former being farther away from the coast. One explanation for the better performance in the central region is that it has more uniform ozone chemistry and diurnal ozone cycle throughout the year than the coastal region, where predominant land-sea breezes may have an impact on ozone chemistry (Figure S4 in the supplementary document shows 24-hour observed ozone concentrations throughout the year. Figures S4-a, b, and c display the three worst-performing stations while Figures S4-d, e, and f display the three best).[12,13] It is evident from the figures that stations with a uniform diurnal ozone cycle provided more accurate forecasts than those with less variability in hourly concentrations. Ideally, the ozone concentration starts to increase afternoon and peaks a few hours before sunset.[9] The CNN model also follows this general ozone chemistry and attempts to make predictions based on this information; hence, the station with generalized ozone chemistry produced more accurate forecasts than the station with less variability in its concentration of ozone throughout the day.

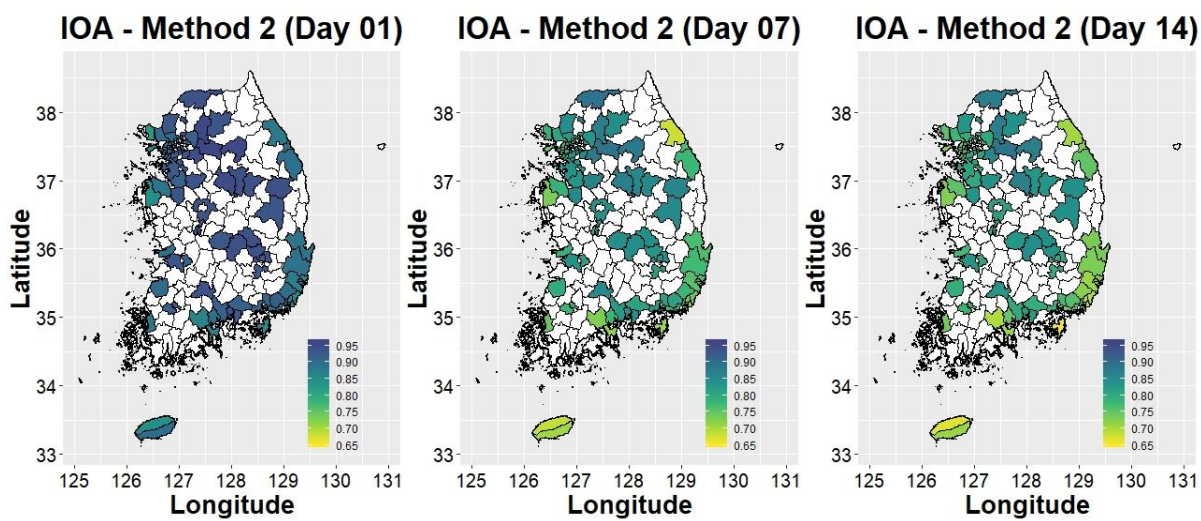

**Figure 2:** Average IOA all stations (CNN-method 2) in each district of South Korea. a) IOA for Day 1; b) IOA for Day 7; and c) IOA for Day 14.

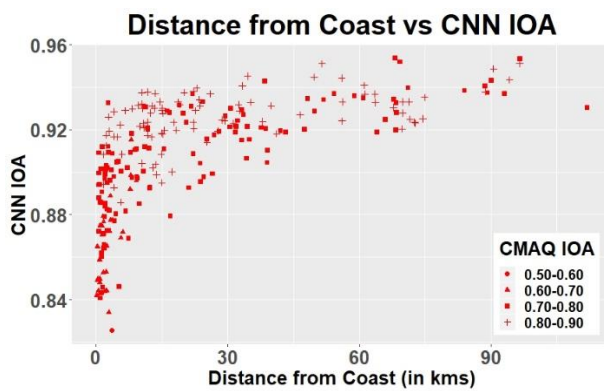

**Figure 3:** Variation of IOA based on distance from the coast. The x-axis represents the distance of the station from the coast, and the y-axis represents the index of agreement. The colored symbols represent the range of CMAQ-IOA for the corresponding station. All IOA are based on one-day ahead prediction only.

Accuracy of forecasting was also highly dependent on the level of urbanization (Figures 2 and S4 in the supplementary document). Out of seven cities with an IOA higher than 0.94, six were among the least urbanized (the 4th and 5th quantiles), and only one was an urban region (the 2nd quantile). Ozone precursors are mostly anthropogenic in urban areas that can be highly variable.[11] This variability leads to a departure from the general diurnal trend of ozone concentrations and thus to the much less accurate forecasting of method 2 in urban areas than in rural areas.

Among the stations on the coastal regions, those on the northwestern coast provided less accurate predictions than those on the northeastern and southeastern coastal cities (Figure S3). A possible explanation for such a trend could be the variability induced by long-range transport from China[14]. The effects of transport are observable at the three stations on Jeju Island (station numbers). Because of transport from the Korean Peninsula, two stations on the northern coast have a lower IOA (0.84 for both stations) than the one station on the southern coast (IOA - 0.90). As a mountain range separates the northern part of the island from the south, transport is blocked.

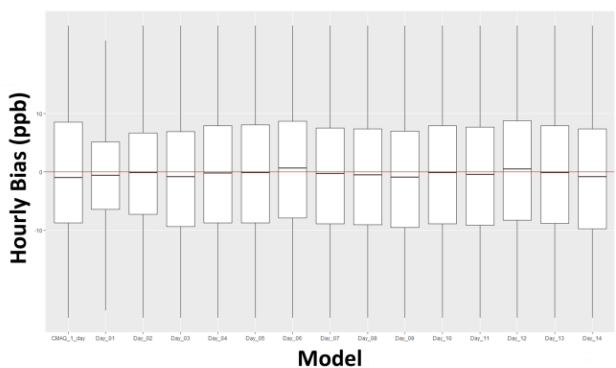

**Figure 4:** Box plot of hourly bias of all stations combined. The x-axis represents the prediction days, and the y-axis represents the hourly bias in ppb. The Redline represents the zero bias, and the black horizontal line in each box represents the mean bias for that model.

Figure 4 shows the boxplot for the hourly bias of all the stations combined for 14-day advance prediction. The bias for one-day advance prediction using the CNN-model is the least. As the number of advance prediction days increases, variability in the bias also increases, but the mean bias remains close to 0 for all days. The day 14 forecast has a similar bias as the one-day advance forecast by the CMAQ model. For all the states, CMAQ overpredicts, except for Busan, Jeju, and Gyeongsangnam-do. CNN's one-day advance prediction shows a slight underprediction (Figures S6 and S7; in the supplementary document). From the second day on, the CNN model initiated overpredictions, which peaked around days 3 and 4 and then began to decrease. Days 7 and 8 showed the fewest overpredictions, and the mean of maximum daily ozone was close to the mean of the observations. The second week followed the same trend as that of the first week. Overprediction increased until the 9$^{th}$ and 10$^{th}$ days, and it decreases. The reason for such weekly trends in the IOA of prediction is that ozone concentrations also followed a weekly trend.[11] Ozone concentrations were strongly auto-correlated with the seventh day, which provided better training of the CNN model for days 7 and 14; hence, the performance of the model on these days improved.

## Conclusion

The predictive accuracy of the CNN model depended on one or a combination of multiple factors: i) the performance of the base model (in this case, CMAQ), ii) distance from the coast, iii) level of urbanization, and iv) transport. These factors, individually or in combination, led to a departure from general diurnal ozone trends. As a result, an anomaly occurred, and in some cases, the model was not able to successfully understand the anomaly, which led to comparatively less forecasting accuracy. The model generally performs better when the CMAQ performs well.

The variability caused by the cyclic reversal of land and sea breeze in ozone concentrations led to poor performance by the CNN model in the coastal region. Distance from that coast has an inverse effect on the prediction accuracy of this CNN model. As we move inland, its accuracy improves. Similarly, as a less urbanized locale has a more consistent diurnal ozone trend, training of the CNN model becomes easier, enhancing its prediction accuracy.

The highly contrasting performance of the model when applied to the western and eastern coasts of South Korea suggest that transport also plays a significant role in determining the accuracy of model predictions. Unlike the eastern coast, the western coast is subject to long-range transport that adds to the variability of ozone trends. This hypothesis was supported by observations of the effects of transport at the three stations on Jeju Island.

The current systems for air quality prediction are either a short-term forecasting system or a low-accuracy system that covers a longer forecasting period. Since this model provides a reasonable forecast two weeks in advance, it can provide an actionable window within which government agencies can deploy effective measures for reducing the occurrence of extreme ozone episodes.

## Methods

The proposed algorithm uses two set of inputs: i) parameters predicted by numerical models and ii) the previous day observed air quality.

### Coupled CMAQ and WRF

To take advantage of numerical modeling, we used air quality and meteorological parameters prepared by the CMAQ v5.2[1] and the Weather Research and Forecasting (WRF) v3.8, covering the eastern part of China, the Korean Peninsula, and Japan, with a 27 km spatial resolution. The detailed configurations of the CMAQ and WRF models are available in Jung et al. (2019).[15]

### Deep Convolutional Neural Network:

We use the deep architecture of the convolution neural network used in Sayeed et al. (2019).[7] The model consists of five convolutional layers and one fully connected layer. We apply convolution to the input features and the elements of the kernel. The final feature map obtained at the end of the first layer of the CNN acts as input for the second layer. Similarly, the output feature map of the second layer is input for the third layer, and so on. In this way, the model has a five-layer CNN, each layer with 32 filters (activation by ReLU), each with a size two kernel randomly initialized by some value for the first iteration. After determining the last feature maps in the last convolutional layer, the fully connected hidden layer with 264 nodes provides the 24-hour output of ozone concentrations. We implemented the algorithm in the Keras environment with a TensorFlow backend.[16,17] (Figure S1 in the supplementary document displays a schematic of the deep CNN architecture for the prediction of hourly ozone concentrations for the next fourteen days.)

A deep CNN, like any neural network, is an optimization problem that attempt to minimize the loss function. The most generally used loss functions are the mean squared error, the mean absolute error, and the mean bias error. In this study, we tested two loss functions: i) the mean square error (method 1) and ii) a customized loss function (method 2) based on the index of agreement (IOA).[18] (The keras function for the IOA as a loss function appears in the supplementary section.) In method 1, the model attempts to find a solution iteratively such that the mean square error is a minimum. Similarly, in method 2, the model attempts to fit it in such a way that the IOA is maximum. In both cases, we obtain two separate models for each day of prediction. The reason for choosing the IOA as a loss function is that high peaked concentrations in air quality forecasting prediction are critical and IOA, unlike the mean bias or the mean square error, is a better parameter that more accurately reports the quality of a model.

**Data Preparation and Model Training**

We obtained observed air quality from the Air Quality Monitoring Stations network, operated by the National Institute of Environmental Research (NIER) for 255 urban stations for the years 2014 to 2017 across the Republic of Korea. The network measures and provides real-time air pollutant concentrations such as sulfur dioxide ($SO_2$), carbon monoxide (CO), ozone ($O_3$) and nitrogen dioxide ($NO_2$). Since the ML model requires continuously measured data for training/testing, we input the missing values of observational datasets. For these missing values, we used SOFT-IMPUTE by Mazumder et al. (2010).[19] We then extracted the concentrations of air pollutants from CMAQ and meteorological parameters from the Meteorology-Chemistry Interface Processor (MCIP) modules of the CMAQ model. For this purpose, we used the temporally and spatially matched CMAQ grid points of the NIER station locations (Table S1 in supplementary document lists all of the parameters extracted from the MCIP and CMAQ).

After acquiring hourly meteorological fields from the WRF model and pollutant concentrations from observations and CMAQ runs, we prepared the input for each station in the form of a two-dimensional matrix in which each column represented a specific parameter (meteorology or gaseous concentration) and each row represented hourly values. Then we trained the model for three years (i.e., 2014 to 2016) and evaluated it for the year 2017. The input dataset consisted of previous 24-hour observed air pollutant concentrations and the following 24-hour forecasted air-pollutants and meteorological field from the CMAQ and WRF, respectively. The output dataset consisted of the next day 24-hour observed ozone concentration. After we defined the inputs and outputs, we combined the datasets from all stations to construct a matrix for training/testing a generalized deep CNN model across the spatial domain. Since we had 255 stations and three years of hourly data for training, we trained the model with 279,480 examples (days), which were further split randomly into equal parts so that the model was trained on one half and validated on the other. Since each parameter had a unique range of values, we normalized each one between zero and 1 to remove the model bias toward any specific high or low valued parameter. It has been observed that having a different maximum and minimum for a training and prediction set destabilizes the model and produces varied results over different runs. Therefore, for the normalization process, we chose "global" maximum and minimum values for each parameter. These global maximum and minimum values guaranteed that none of the hourly values exceeded a certain level; thus, the normalization process remained independent of the temporal and spatial variations. After normalization, we used the deep CNN architecture (defined in the previous section) to train the model and generated two models, each with a unique loss function. Once the model was generated, it was used to predict the entire year of 2017.

For long-term training and prediction, we prepared the dataset so that it had the same inputs, but we changed the outputs from the first day to the second, third, and fourth days and so on until the fourteenth day (Figure S2 in the supplementary document presents a schematic diagram of the data setup used in this study.) Hence, with two loss functions and 14 days of predictions, we had 28 models to evaluate.

## Acknowledgments

This study was supported by the High Priority Area Research Seed Grant of the University of Houston.

## Data availability

The test/train/validation data in are available for non-commercial research purpose by contacting the corresponding author.

## Code availability

Code for the algorithm development, evaluation and statistical analysis is freely available for non-commercial research purposes by contacting the corresponding author.

# Supplemantary Document

**Supplementary Figures:**

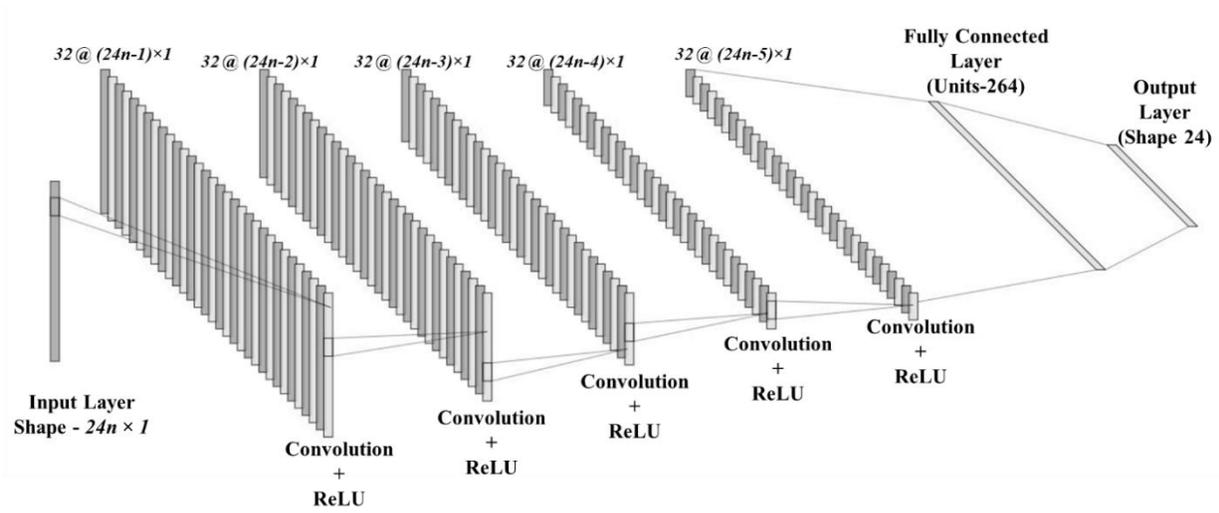

**Figure S1:** Schematic diagram of the Convolution Neural Network. 'n' is the number of input parameters (meteorology, air quality, and observations) used.

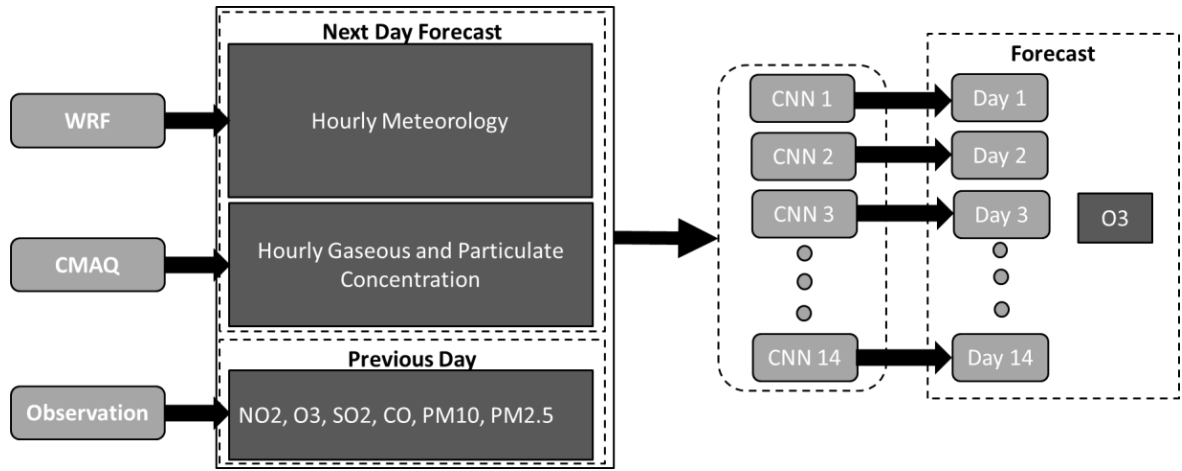

**Figure S2:** Schematic diagram of the process flow of the CNN model.

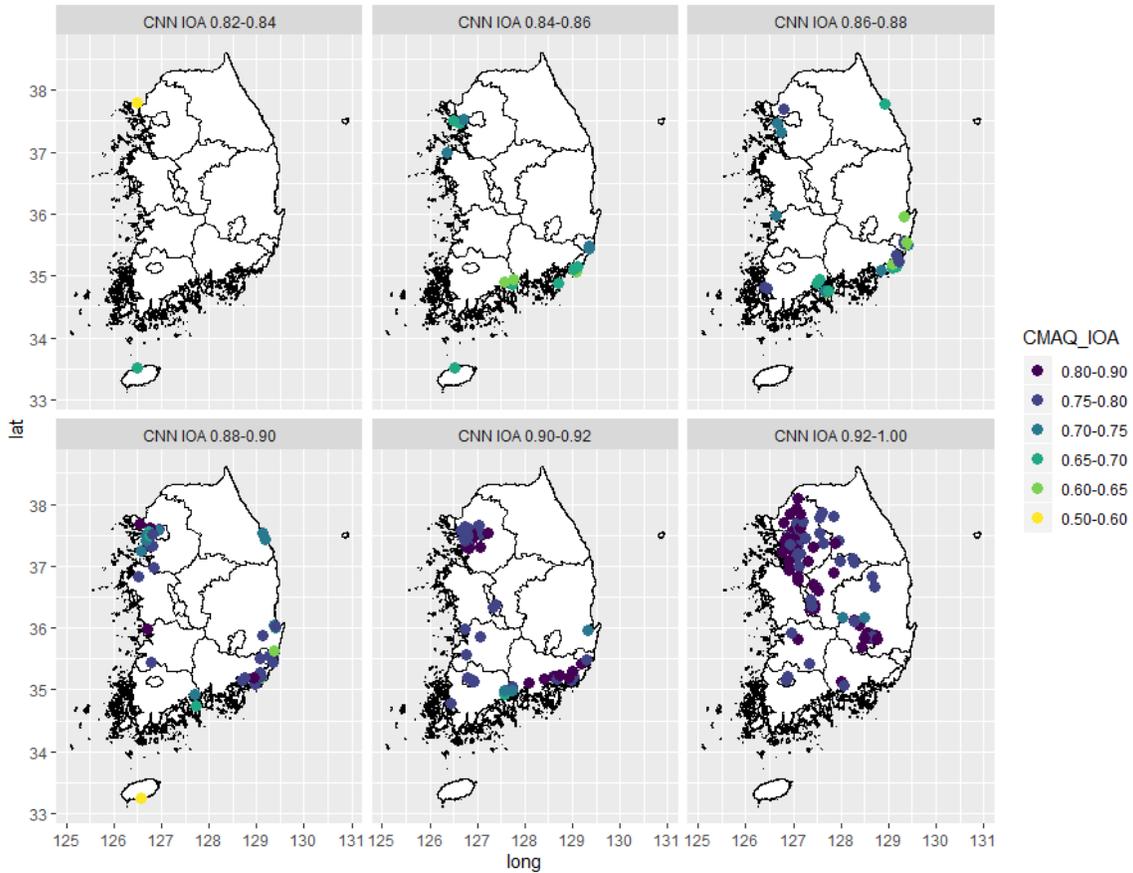

**Figure S3:** Station-based CNN-IOA binned in specific ranges. A colored dot represents the location of the station, and a specific color represents the CMAQ-IOA.

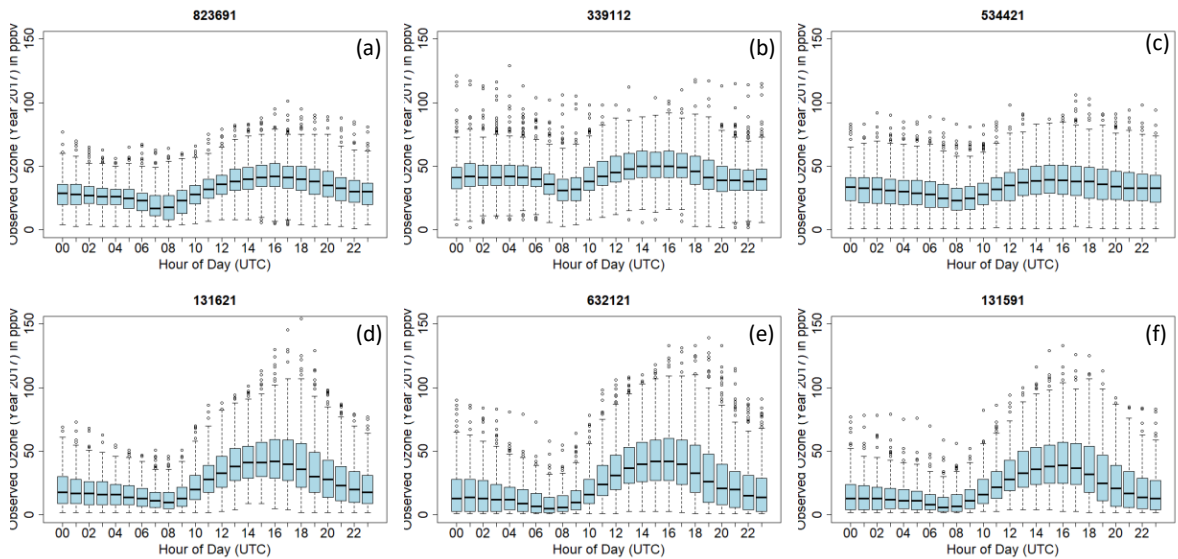

**Figure S4:** Box and whisker plot 24-hour observed ozone concentration throughout the year 2017. a, b and c are three worst-performing stations. d, e, and f are the best performing station.

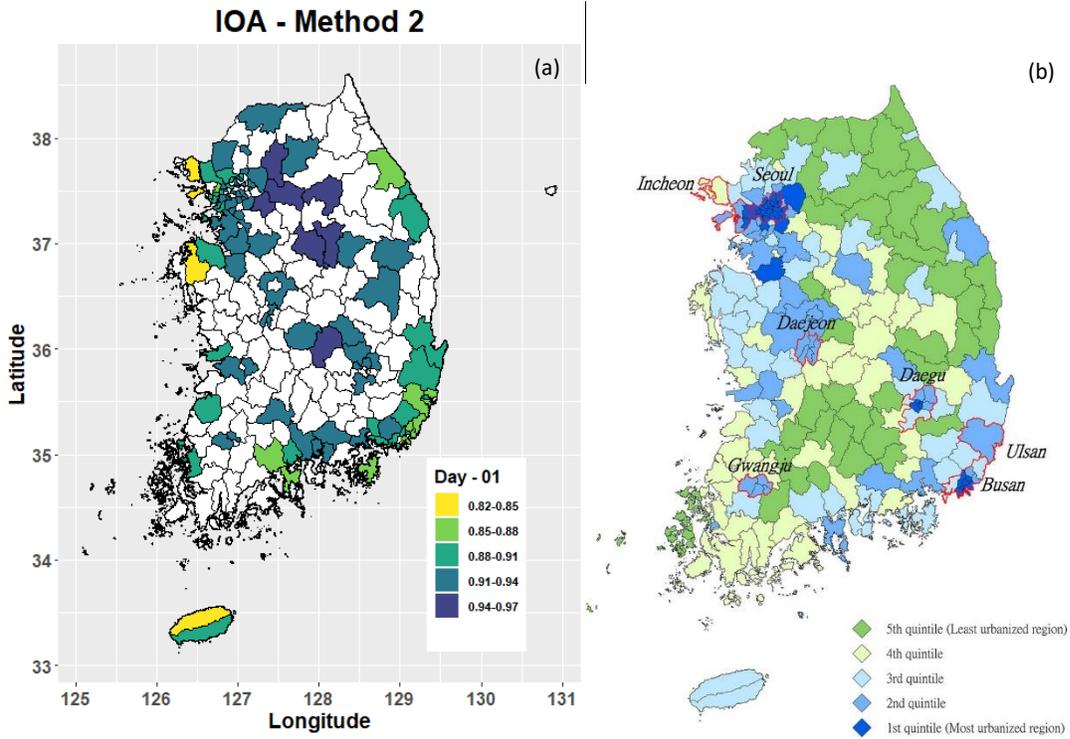

**Figure S5:** a) District-wise IOA based on Method 2 of CNN. b) Level of urbanization in each district (Image Source: Chan et al., 2015)[1]

**References:**
1. Chan, C. H., Caine, E. D., You, S. & Yip, P. S. F. Changes in South Korean urbanicity and

    suicide rates, 1992 to 2012. *BMJ Open* **5**, e009451 (2015).

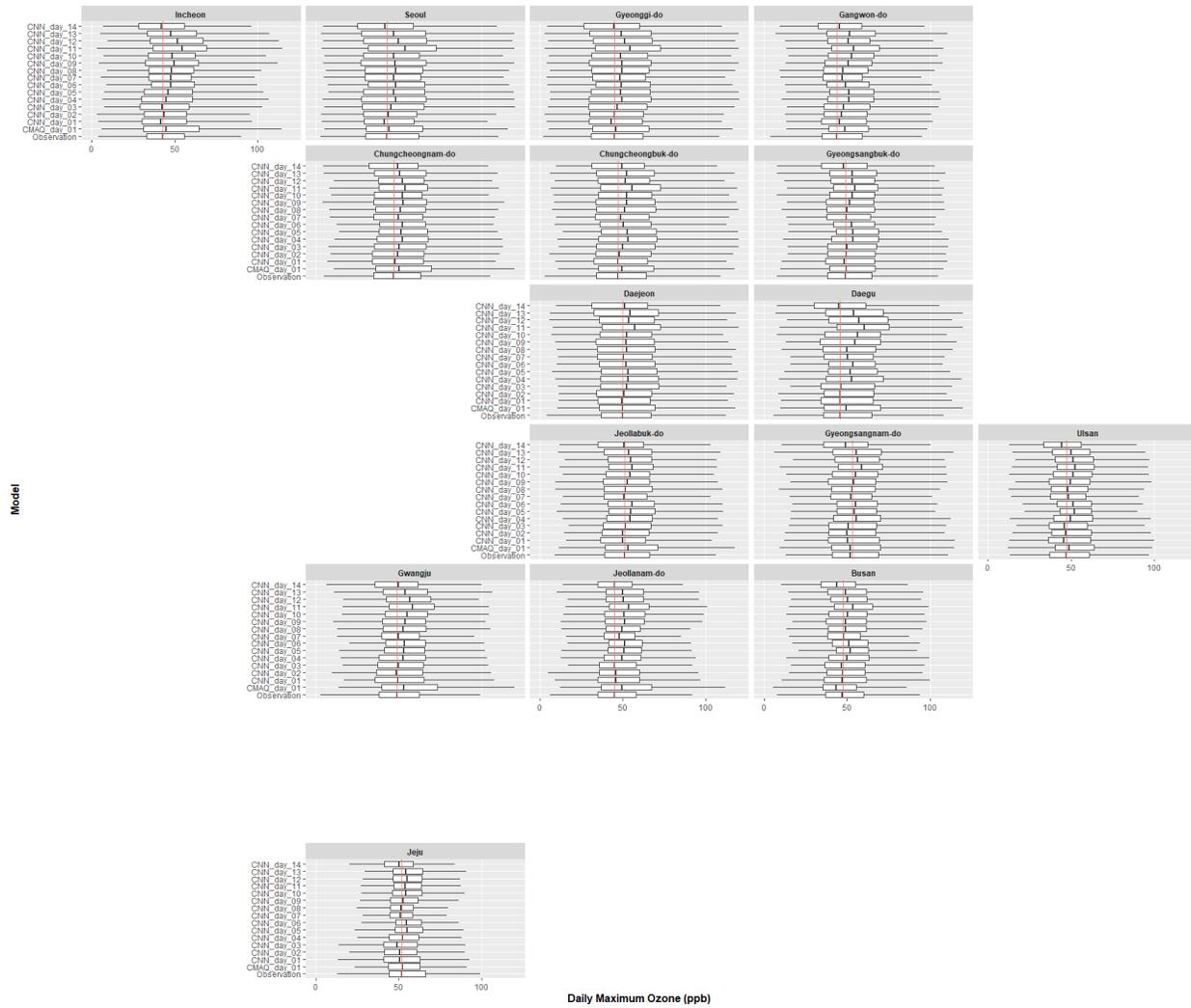

**Figure S6:** State-wise box plot for a daily maximum of ozone concentration for observation, CMAQ prediction, and all fourteen days prediction by CNN. The x-axis represents the Daily Maximum of ozone concentration, and the y-axis represents the prediction day.

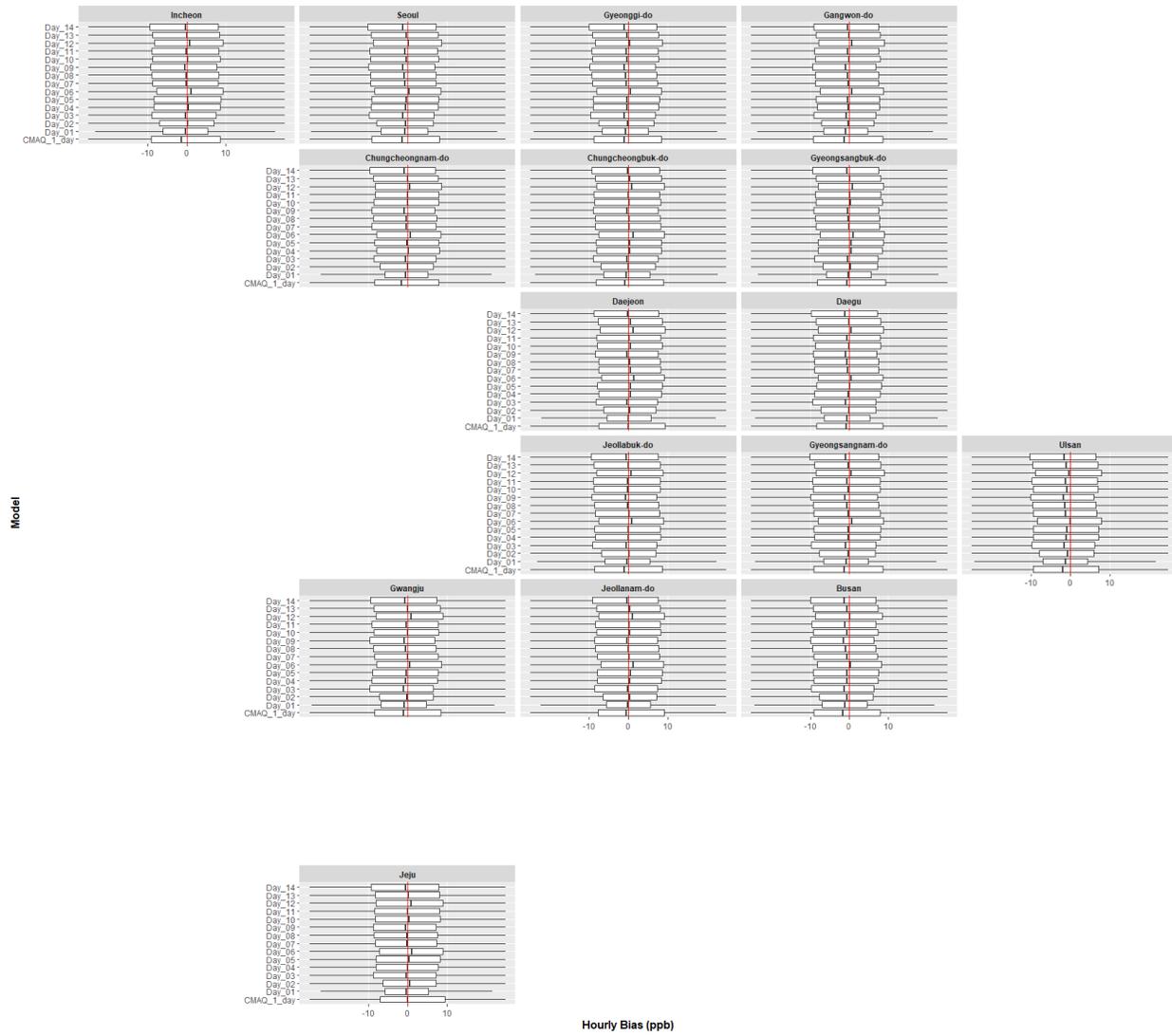

**Figure S7:** State-wise box plot for hourly bias in ozone concentration for CMAQ prediction and all fourteen days prediction by CNN. The x-axis represents the Daily Maximum of ozone concentration, and the y-axis represents the prediction day.

**Supplementary Table:**

**Table S1:** List of parameters from WRF/MCIP and CMAQ used to train the CNN model.

| Abb. | Variable Name (WRF/MCIP) | Units |
|---|---|---|
| **PRSFC** | Surface Pressure | Pascal |
| **USTAR** | Cell Averaged Friction Velocity | m/s |
| **WSTAR** | Convective Velocity Scale | m/s |
| **PBL** | Planetary Boundary Level Height | M |
| **MOLI** | Inverse Of Monin-Onukhov Length | 1/m |
| **HFX** | Sensible Heat Flux | watt/m$^2$ |
| **RADYNI** | Inverse Of Aerodynamic Resistance | m/s |
| **RSTOMI** | Inverse Of Bulk Stomatal Resistance | m/s |
| **TEMPG** | Skin Temperature At Ground | Kelvin |
| **TEMP2** | Temperature At 2 M | Kelvin |
| **Q2** | Mixing Ratio At 2 M | Kg/Kg |
| **WSPD10** | Wind Speed At 10 M | m/s |
| **WDIR10** | Wind Direction At 10 M | Degrees |
| **GLW** | Longwave Radiation At Ground | watt/m$^2$ |
| **GSW** | Solar Radiation Absorbed At Ground | watt/m$^2$ |
| **RGRND** | Solar Rad Reaching Sfc | watt/m$^3$ |
| **RN** | Nonconvec. Pcpn Per Met Tstep | cm |
| **RC** | Convective Pcpn Per Met TSTEP | cm |
| **CFRAC** | Total Cloud Fraction | fraction |
| **CLDT** | Cloud Top Layer Height (M) | meter |
| **CLDB** | Cloud Bottom Layer Height (M) | meter |
| **WBAR** | Avg. Liquid Water Content Of Cloud | g/m$^3$ |
| **SNOCOV** | Snow Cover | fraction |
| **VEG** | Vegetation Coverage (Decimal) | Fraction |
| **LAI** | Leaf-Area Index | m$^2$/m$^2$ |
| **SEAICE** | Sea Ice | Fraction |
| **WR** | Canopy Moisture Content | M |
| **SOIM1** | Volumetric Soil Moisture In Top Cm | m$^3$/m$^3$ |
| **SOIM2** | Volumetric Soil Moisture In Top M | m$^3$/m$^4$ |
| **SOIT1** | Soil Temperature In Top Cm | Kelvin |
| **SOIT2** | Soil Temperature In Top M | Kelvin |
| **SLTYP** | Soil Texture Type By USDA | Category |

| Abb. | Variable Name (CMAQ) | Units |
|---|---|---|
| **NO2** | Nitrogen Di Oxide | ppmV |
| **NO** | Nitric Oxide | ppmV |
| **O** | Oxygen Atom | ppmV |
| **O3** | Ozone | ppmV |
| **NO3** | Nitrate | ppmV |
| **O1D** | Oxygen Atom | ppmV |
| **OH** | Hydroxide | ppmV |
| **HO2** | Hydroperoxyl | ppmV |
| **N2O5** | Nitrogen Pentoxide | ppmV |
| **HNO3** | Nitric Acid | ppmV |
| **HONO** | Nitrous Acid | ppmV |
| **H2O2** | Hydrogen Peroxide | ppmV |
| **CO** | Carbon Monoxide | ppmV |
| **PAN** | Peroxyacyl Nitrates | ppmV |